\nopagenumbers
\def \half {{\textstyle {1\over 2}}}
\def \twothird {{\textstyle {2\over 3}}} \def \a {{\alpha }} \def \m {\mu}
\def \n {\nu}
\def \Tr {\rm Tr}

\def \e {{\epsilon}}
\magnification =\magstep 1
\overfullrule=0pt
\hsize = 6.5truein
\vsize = 8.5truein
\null
\vskip-1truecm
\hfill{IC/98/127}
\vskip.1truecm
\hfill{hep-th/9809044}
\vskip.5truecm
\centerline{United Nations Educational Scientific and Cultural Organization}
\centerline{and}
\centerline{International Atomic Energy Agency}
\medskip
\centerline{INTERNATIONAL CENTRE FOR THEORETICAL PHYSICS}
\vskip1.5truecm
\centerline{{\bf Aspects of Six Dimensional Supersymmetric
Theories}\footnote{$^\dagger$}{Contribution to the Abdus Salam Memorial
Meeting, 19-22 November 1997, Trieste, Italy. }}
\vskip 2cm
\centerline{ S. Randjbar--Daemi\footnote{$^{\dagger\dagger}$}{e-mail
address: {\tt daemi@ictp.trieste.it}}}
\vskip .1 in
\centerline{International Centre for Theoretical Physics, Trieste, Italy}
\vskip 1.5cm
\centerline{ABSTRACT}
\baselineskip=18pt
\bigskip
{In this contribution some aspects of supergravity and super
Yang-Mills systems in $D=6$ are briefly reviewed and, in some cases,
are contrasted with the analogous features in $D=4$. Particular
emphasis is laid on the stringy solutions of the $D=6$ super
Yang-Mills systems.}
\vskip 1.5cm
\centerline{MIRAMARE -- TRIESTE}
\centerline{May 1998}
\vfill\eject
\footline{\hss\tenrm\folio\hss}
\pageno=2
\noindent{\bf 1. Introduction}.

I was indeed very privileged to be for many years a close associate of
Abdus Salam. I have learned many things from him.  Salam combined the
vigorous western thought in a coherent manner with his oriental
culture. He believed deeply that the social life of an individual has
a sense and purpose only in relationship with those of others. It
becomes richer and more purposeful if its guiding principles are
compassion and tolerance. He himself was a proverbially generous
person.

Salam deeply appreciated the relevance of Science to the enrichment of
human life. He spent a major part of his active life to disseminate
scientific knowledge among the less privileged nations.  Being a
singularity as he was, he also contributed substantially to the
advancement of the fundamental science. In fact the very best existing
theory of Nature, the Standard Model of Particle Physics, bears his
name.

My scientific collaboration with Abdus Salam started with a study of
theories of Kaluza - Klein type in a space time of six dimensions
[1]. I have therefore chosen to review in this memorial contribution
some of the recent developments in 6-dimensional theories. The
presentation will be mostly, but not always, non technical and
elementary.

\noindent{\bf 1. Particles and Strings in D=4 and D=6}

Physical theories in a six dimensional manifold of Lorenzian signature
differ in many respects from the four dimensional theories. In $D=6$
in addition to spinors, scalars, vectors, and second rank symmetric
tensors, which are the basic objects of $4$-dimensional field
theories, we also have second rank antisymmetric tensor
potentials. Also the fact that the fundamental spinor representation
of $SO(1, 5)$ is pseudo real, as opposed to the complex Weyl spinors
of $SO(1,3)$, has some significance in constructing anomaly free
models in D=6. We shall start with a summary of differential forms and
the extended objects to which they couple [2].

In $D=4$ the only interesting forms are the 1-forms and their exterior
derivatives which
correspond to, respectively, a Maxwell (or Yang-Mills) potential $A$ and
its field strength $F$.
Being a $2$ form, $F$ admits a dual, $^{*}F$, which is
also a $2$ form. Maxwell's equations are essentially symmetric under the
exchange of $F$ and
$^{*}F$. This is called the
electromagnetic duality under which the electric and magnetic charges
interchange their roles [3].
This type of duality is the prototype of a larger class of duality
symmetries which can occur in
space times of higher than 4 dimensions. Note that in $D$ dimensions the dual
of a $p$ form is a $D-p$ form. It is thus only for $D=4$ that the dual of
the electromagnetic $2$
form $F$ is again a $2$ form. To appreciate the physical significance of
this simple fact let us
recall
that the electromagnetic potential $A$ couples to particles through the
term $\int_C A$ where $C$ denotes the world line of the charged particle.If
$^{*}
F$ is derived from a dual potential $\tilde A$ then there will be a dual
particle which could couple to $\tilde A$ through $\int_{\tilde C}\tilde
A$, where $\tilde C$ is
the world line of the dual particle. In $D$ dimensions a $p+1$ form
potential couples naturally to a
$p$ dimensional extended object, called a p-brane. This coupling is a
direct generalization of
the electromagnetic coupling, namely, $\int_{\Sigma_{p+1}} A$, where
$\Sigma_{p+1}$ denotes
the
$p+1$ dimensional world volume of the $p$ dimensional extended object. Note
that a
$p$ brane will occupy a $p$ dimensional subspace of the $D$ dimensional
space time. For
example, we can think of a $p$ dimensional hyperplane extended along $p$ of
the $D-1$ space
coordinates. Therefore, in the remaining $D-1-p$ space dimensions our
object will look like a
point, which we call the position of the $p$ brane. A large sphere around
this position will have
$D-2-p$ dimensions, which equals to the rank of the dual of the $p+1$ form
$F=dA$. We can
thus integrate $^{*}F$ over this large sphere and call it the electric
charge of the $p$ brane. On the other hand the integral of the $p+2$ form
$F$ over a $p+2$
dimensional sphere is called the magnetic charge of the considered $p$
brane. These are direct
extensions of the well known definitions in $D=4$. Thus in $D$ space time
dimensions the dual of
a $p$ brane is an extended object with $D-p-4$ space dimensions, i.e. a
$D-p-4$ brane.

With the above definitions, in $D=6$ the dual of a particle is a $2$ brane,
while
the dual of a $1$ brane is again a $1$ brane. This indicates that in $D=6$
one dimensionally
extended objects, namely, strings, and the two form potentials to which
they couple, play a role
analogous to
the role of particles and vector potentials in $D=4$.

To mention other
interesting differences let us impose the condition $F= ^{*}F$. This
condition is
meaningful only for a $2$ form $F$ in $D=4$ and a $3$ form $F$ in $D=6$.
Thus in $D=4$, if
the self-duality equation had a real solution, there could exist self dual
particles. However, it turns
out that the self duality equation in $D=4$ has interesting solutions only
if the $4$ dimensional
manifold is Euclidean and the gauge group is non Abelian. These solutions
are called instantons.
They are localised finite action solutions of the $4$ dimensional Euclidean
Yang-Mills equations.
The value of their Euclidean action, appropriately normalised, equals to
their topological charge.
These solutions do not have particle type interpretation in $4$ dimensional
Minkowski space time.
In contrast with $D=4$,
the self duality equation $F= ^{*}F$
has a solution for an Abelian
$3$ form $F$ in $D=6$ Minkowski space-time. These are self-dual strings
[4]. Such strings will
carry both
electric and magnetic charges and their magnitude will be equal. The non
Abelian version of
higher rank forms are not yet known.

One may wonder what type of interpretation the ordinary Yang-Mills
instantons can have in
$D=6$. To answer this question we need to consider a non Abelian gauge
theory in a
6-dimensional space time. One can consider a $4$-dimensional Euclidean
subspace and an
instanton configuration localised in this subspace.
{}From the point of view of the 6-dimensional space time this
object will look like a string with a thickness given by the size of the
instanton.
As the size goes to zero it will look more and more like a fundamental
string[5].
We shall make use of this interpretation in section three.

\noindent{\bf 2. Supersymmetry and Chiral Anomalies}

The possibility of having self dual or anti-self dual gauge fields in $D=6$
make the
six dimensional supersymmetric theories more akin to D=10 theories rather
than their D=4
counterparts. The conditions for the cancellation of chiral anomalies are
also more stringent in
$D=6$. For example, there are no pure gravitational
anomalies
in $D=4$ theories. Such anomalies can exist, however, in $D=6$. The
requirement that the pure
gravitational anomalies do cancel imposes restrictions on the
supermultiplet structure
of the six dimensional supergravity theories.

A model with a minimum number of supersymmetries in $D=6$ has four
independent complex
supercharges
which can be assembled into a Weyl spinor of $SO(1,5)$. It is customary to
double the number of
components and impose a symplectic Majorana condition. For this reason such
models are
sometimes called $(2,0)$. We shall refer to them as $(1,0)$. These
models like their extended versions with $16$ real supercharges which we
shall denote as $(2,0)$
are chiral, while their $(1,1)$ and $(2,2)$ theories, which have
respectively $16$ and $32$
supercharges,
are non chiral.

In section 5 we shall give a little more detail about the models with more
than $8$ real
supercharges
and explain briefly how they can be obtained from the $D=10$ superstring
theories upon
compactification on a
four manifold. In the rest of this section we shall exclusively discuss
theories with $(1,0)$
supersymmetry.

The $(1,0)$ models admit the following type of super multiplets:

1) gravity: $E_a^m, \Psi_\mu, B_{ab}^{-}$

2) hypermatter : $\psi^r, \phi^\alpha$

3) Yang-Mills: $\lambda, A_a, Y $

4) Tensor : $\chi, B_{ab}^{+}, \sigma$

The spinors in the gravity and the Yang-Mills multiplets are left handed,
while those
in the tensor and hypermultiplets are right handed with respect to
$SO(1,5)$. Furthermore, the
spinors in the gravity, Yang-Mills and the tensor multiplets are doublets
of an automorphism
$Sp(1)=SU(2)$, and they
are Majorana symplectic in the sense that they satisfy a constraint of the
type $\psi=
\Omega\psi_{c}$ where $\Omega$ is the $Sp(1)$ invariant metric and
$\psi_{c}$ is the charge
conjugate of $\psi$. The superscripts $\pm$ on the antisymmetric tensor
potentials indicate that
their field strengths
are self dual (+) or anti-self dual (-). The scalars in the hypermultiplets
are the $2n$ complex
coordinates of a quaternionic manifold. Thus $\alpha=1,...,2n$ and the index
$r$ in the hypermatter fermions, $r=1,...,n$, where, $n$ counts the number
of the hypermatter
multiplets.
$Y$ in the Yang-Mills multiplet is an auxiliary field. It is a triplet of
the automorphism $Sp(1)$.

In any $D=6$ coupled Yang-Mills supergravity theory, the condition for the
cancellation of pure
gravitational anomalies imposes a restriction on the number of multiplets
listed above. This
condition is[6]
$$ n=m+273-29k$$
where $m$, $n$ and $k$ are respectively the number of Yang-Mills, hyper and
tensor multiplets.

Anomaly free supergravity models with $(1,0)$ supersymmetry in $D=6$ can be
obtained from
the
compactifications of the ten dimensional heterotic strings on a particular
class of complex
manifolds called $K_3$ [7].

Having cancelled the pure gravitational anomalies the remaining anomalies
can be cancelled with
the help of a mechanism discovered by Green and Schwarz in the context of
ten dimensional
superstring models [8] and extended to the six dimensional models in [6, 7].

Let us consider models with $k=1$. For such models one can construct
invariant supergravity
actions. The anomaly condition simplifies to $n=m+244$. The Green-Schwarz
anomaly cancellation mechanism starts from an $8$ form $P_8=X_4.\tilde
X_4$, where "."
indicates a wedge product. The $4$ forms $X_4$ and $\tilde X_4$ have the
following general
structures $$ X_4= tr R^2- \Sigma v_\alpha trF_\alpha^2$$ $$\tilde X_4= tr
R^2- \Sigma\tilde
v_\alpha trF_\alpha^2$$ where $v_\alpha$ and $\tilde v_\alpha$ are
numerical constants and
$F_\alpha$ is the field strength $2$ form associated with the gauge group
$G_\alpha$. Note that
both $X_4$ and $\tilde X_4$ are closed. Locally we can write $X_4= d
\Omega_3$, where
$\Omega_3$ is a Chern-Simons three form. It is not gauge invariant. Since
$X_4$ is gauge
invariant, under a gauge transformation, we need to have $\Omega_3
\rightarrow \Omega_3 +
d\alpha$. The $2$ -form $\alpha$
can be constructed explicitly [9]. Up to total derivatives, the gauge and
gravitational chiral anomaly
is proportional to the integral of the $6$ form $\alpha.\tilde X_4$ over
the Euclideanised
six-dimensional space time.

The GS mechanism requires the
addition of a local counter term of the form $B.\tilde X_4$ to the
effective Lagrangian. If we
demand that the two form potential $B$ undergoes a gauge transformation of
the form $B
\rightarrow B- \alpha$ then the one loop effective action will be gauge
invariant. A gauge invariant
field strength associated with the $2$ form potential $B$ should thus be
defined by $ H=dB +
\Omega_3$.

All this is very similar to the application of the Green-Schwarz mechanism
to the anomaly problem
in $D=10$ heterotic string theory. In that case the anomaly polynomial is a
$12$ form which
factorises as $P_12=X_4.X_8$, where $X_4$ and $ X_8$ are
closed forms and furthermore $X_4= {1\over {30}} TrF^2 - tr R^2$. Here $Tr$
and $tr$ refer,
respectively, to the adjoint and the fundamental representations.

The modified definition of $H$ has very interesting consequences for $K_3$
compactifications.
First note that $$dH= tr R^2- \Sigma v_\alpha trF_\alpha^2$$ Integrate this
expression on $K_3$.
If $dH$ has no $\delta$-function type singularities on $K_3$ its integral
will vanish. We then
obtain $24= \int \Sigma v_\alpha trF_\alpha^2$, where $24$ is the Euler
number of $K_3$.
The integral on the right-hand side of this expression is the Chern number
(instanton number)
associated with the background gauge field configuration. For the
$E_8\times E_8$ heterotic string
it will equal to the sum of instanton numbers embedded in each $E_8$
factor. We thus obtain
$24=n_1+n_2$.

If $dH$ has singularities they can be interpreted as $5$ branes and the
result of integration
becomes $24=n_1+n_2 +n_5$ for the $E_8\times E_8$ and $24=n+n_5$ for the
$SO(32)$
heterotic strings, where $n_5$ indicates the number of $5$ branes.

In the above discussion we considered only models with one tensor
multiplet. A more general
framework has been developed in [10] for models with $k>1$

In addition to anomalies in local symmetries there is also the possibility
of global anomalies in the
supergravity and super Yang-Mills theories in $D=6$. Such anomalies exist
in $D=4$ theories if
the fourth homotopy group, $\Pi_{4}$, of the gauge group is non trivial
[11]. If $\Pi_{6}$ of
the gauge group is non trivial, a super Yang-Mills theory in $D=6$ can be
inconsistent due to the
presence of global gauge anomalies. This happens for the groups $SU(2)$,
$SU(3)$ and $G_2$.
The requirement of
the absence of such anomalies imposes further restrictions on the structure
of the consistent six
dimensional models. We shall return to this issue in section 4.

Very interesting physics can be extracted by studying the moduli space of
scalars.
We saw above that there are two types of scalars in the spectrum of $(1,0)$
models, namely, those
in the hypermultiplets and those in the tensor multiplets. Seiberg and
Witten parametrise the
Coulomb branch of the $(1,0)$ theories by the expectation
values of the scalars in the tensor multiplets [12]. The reason for naming
the tensor moduli space
as the Coulomb branch is that upon compactification to lower dimensions
the tensor multiplets become vector multiplets of the lower dimensional
theories.

The metric in the hypermultiplet moduli space is independent of the scalars
in the tensor multiplets.
Likewise the metric in the Coulomb branch and the kinetic energy terms for
the vector and the
tensor fields are independent of the scalars in the hypermultiplets. In the
infrared
region, where the physics is described by classical field theory, there is
no supersymmetric
coupling which could lead to masses for massless particles by wondering in
the moduli space of
scalars in the tensor multiplets.
Conversely, it is impossible for massive particles to become massless at a
particular point on the
Coulomb branch, by any mechanism
that can be described at low energies by a free field theory. Seiberg and
Witten then
conclude that the singularities in the Coulomb branch necessarily involve
non infrared freephysics
which is associated with non-critical tensionless strings, where non-critical
means that gravity is not a mode of the string. Thus the non trivial
infrared physics
associated with the Coulomb branch singularities is occurring in flat six
dimensional
Minkowski space time. In the next section we shall study a flat space model
which exhibits some
of these features.

\noindent{\bf 3. Super Yang-Mills coupled to tensors in D=6 and non
critical Strings}

As it was argued in the preceding section the singularities in the Coulomb
branch of $(1,0)$
models are conjectured to be related to non critical strings which do not
couple to gravity and
therefore they can be studied in a flat Minkowski space time.
Supersymmetric models in flat
$D=6$
which involve only the Yang-Mills and the tensor multiplets have been
constructed in [13]. In
addition to the argument
given at the end of the last section, the introduction of tensor multiplets
is called for also by the
requirement of anomaly cancellation, because, it is the presence of such
tensor fields in six
dimensions which enable us to make use of the
Green-Schwarz anomaly cancellation mechanism. The tensor multiplet by
itself does not have
nonsingular solitonic solutions. This is one more reason for considering
the tensor multiplet
coupled to Yang-Mills fields. \footnote *{This section
and section $4$ follow closely ref.[14]. Similar results have been obtained
from the flat space limit
of a $D=6$ supergravity model in [15].}

The supersymmetry transformations of the fields in our system are as
follows [13].
$$\eqalign{
\delta A_a&= -\bar\epsilon_i~\Gamma_a\lambda^i\cr \delta \lambda^i &=
{1\over 8}\Gamma^{ab}
F_{ab} \epsilon^i - {1\over 2} Y^{ij}\epsilon_j\cr \delta Y^{ij} &= -
{\bar\epsilon}^{(i}~\Gamma^a D_a\lambda^{j)}\cr} \eqno(1) $$ where the
index $i$ is a doublet
index of the automorphism $Sp(1)$.
The corresponding rules for the tensor multiplet coupled to Yang--Mills are
given by
$$\eqalign{
\delta \sigma &= \bar\epsilon\chi\cr
\delta\chi^i &= {1\over
48}\Gamma^{abc} {H}^{+}_{abc}
\epsilon^i +{1\over 4}\Gamma^a\partial_a\sigma\epsilon^i - {\a'\over 4}{\rm
Tr}\,\Gamma^a\lambda^i
\bar\epsilon\Gamma_a\lambda \cr
\delta B_{ab} &= -\bar\epsilon\Gamma_{ab}\chi - \a' {\rm Tr}\,
A_{[a}\bar\epsilon
\Gamma_{b]}\lambda
\cr}\eqno(2)
$$
where
$$\eqalignno{
{ H}_{abc} &= 3\partial_{[a} B_{bc]} + 3\a'{\rm Tr} \left(A_{[a}\partial_b
A_{c]}+{\twothird}
A_a A_b A_c\right)\cr { H}_{abc}^{\pm} &= {1\over 2}\biggl ( { H}_{abc} \pm
\tilde {
H}_{abc}\biggr ).&(3)\cr}
$$

The closure of the supersymmetry algebra leads to the following field
equations for various fields,
$$\eqalignno{
{ H}_{abc}^{-} &= -{\a'\over 2}{\rm Tr}\,( \bar\lambda
\Gamma_{abc}\lambda)&(4a)\cr
\Gamma^a\partial_a\chi^i &= \a' {\rm Tr}\,\left({1\over 4}\Gamma^{ab}
F_{ab} \lambda^i +
Y^{ij}\lambda_j\right)&(4b)\cr \partial^2 \sigma &= \a' {\rm
Tr}\,\left(-{1\over 4}F^{ab}F_{ab}
- 2\bar\lambda \Gamma^a D_a\lambda + Y^{ij}Y_{ij}\right).&(4c)\cr }$$
Further, by virtue of its
definition,
$H_{abc}$ satisfies the identity
$$
\partial_{[a}{ H}_{bcd]}^+ = \a' {\rm tr}\,\left({3\over 4} F_{[ab}F_{cd]} -
\bar\lambda\Gamma_{[abc}D_{d]}\lambda\right) \eqno(5) $$

We shall look for a bosonic background configuration in which all the
fermions as well as the
auxiliary
field $Y^{ij}$ will vanish. The six-dimensional coordinates will be chosen
as $x^+$, $x^-$ and
$x^{\m}$ where $\m =1,...4$. We shall consider a multi instanton-type
configuration in the ${\bf
R}^4$ spanned by $x^{\m}$. It will be shown that that the moduli of this
instanton can depend on
$x^+$. This will require that the $A_+$-component of the vector potential
is different from zero.
In this sense the solution looks like a static monopole configuration in
the six-dimensional
spacetime in which $x^-$ is taken to be the time coordinate. This
configuration will
preserve half the six-dimensional $N=1$ supersymmetry.

It follows from (4a) that if $\lambda =0$, then $H$ is self dual.\footnote
{*}{ We shall henceforth
drop the superscript $+$ from H.}. Now setting $\delta \lambda$ and
$\delta\chi$ equal to zero we
obtain
$$
\Gamma^{ab} F_{ab} \epsilon =0,~~~~~~~~~ \left( \Gamma^a\partial_a \sigma
+{1\over 12}
\Gamma^{abc}H_{abc} \right ) \epsilon =0 \eqno(6) $$
To satisfy these equations, we can choose
$\epsilon=\left(\matrix{\varepsilon\cr 0}\right)$, where,
$\gamma_5 \varepsilon =\pm
\varepsilon$, and
$\gamma_5$ gives the four-dimensional chirality. We shall first discuss the
case
of positive chirality; the case of negative chirality can be obtained by
essentially
trivial change of some self duality conditions. With this choice the fields
must obey the equations
$$
\eqalignno{
H_{05\m} &= -\partial_\m \sigma &(7a)\cr H_{0\m \n} &= {\tilde H_{0\m\n}}=
\half \e_{\m\n
\alpha \beta}H_{0\alpha \beta}&(7b)\cr F_{\m\n}&= {\tilde F_{\m\n}}= \half
\e_{\m \n \alpha
\beta}F_{\alpha \beta}&(7c)\cr } $$ together with $F_{+-}=0, ~F_{-\m}=0,
~\partial_- \sigma
=0$. Choosing the gauge $A_- =0$,
these reduce to
$\partial_- A_\m = \partial_- A_+ = \partial_- \sigma =0$.

The constraint (5) for $H_{abc}$, expressing its coupling to the Yang-Mills
fields via the
Chern-Simons 3-form, now gives the following conditions, $$\eqalignno{
\partial_- H_{+\m\n}&=0&(8a)\cr
\partial_\lambda H_{+\lambda \alpha} &= \partial_+ \partial_\alpha \sigma ~
-2c \Tr (F_{\lambda
\alpha}{ F}_{+\lambda})&(8b)\cr \partial_\m \partial_\m \sigma &= -{c\over
2} \Tr (F_{\m\n}
{\tilde F}_{\m\n})&(8c)\cr } $$
where $c= 3 \a '/4$.
Further, $H_{-\m\n}=0$ and $H_{+\m\n}= {\tilde H}_{+\m\n}$. Setting the
auxiliary field
$Y^{ij}$ to zero implies $D_a F^{ab}=0$ [13]. The only nontrivial surviving
component of this
equation is $$
D_\lambda (D_\lambda A_+ - \partial_+ A_\lambda )=0 \eqno(9) $$ where
$D_\lambda A_+ =
\partial_\lambda A_+ + [A_\lambda ,A_+]$.

The strategy for solving these equations is as follows. We first choose
$F_{\m\n}$ to be a
multi-instanton configuration in ${\bf R}^4$. Then equation (8c) gives
$\sigma$, and (7a) gives
$H_{05\m}$. Since $D_\lambda D_\lambda$ is invertible in the instanton
background, (9) can be
uniquely solved for $A_+$. Finally, equation (8b) can be solved,
consistently with its self duality,
to get $H_{+\m\n}$.
As a consequence of $\partial_- A_\m = \partial_- A_+ =0$, the instanton
parameters,
collectively denoted by $\xi$, obey
the condition $\partial_- \xi =0$, but they can, of course, depend on
$x^+$. (They are thus
left-moving modes in the $(x^0,x^5)$-subspace.)

Using the self duality of $H_{+\m\n}$, we can rewrite (8b) as $$
\partial_\lambda \partial_\lambda
H_{+\m\n} = (\partial_\m J_\n -\partial_\n J_\m ) +\half \epsilon_{\m\n
\alpha \beta}
(\partial_\alpha J_\beta -\partial_\beta J_\alpha ) \eqno(10) $$ where
$J_\alpha = \partial_+
(\partial_\alpha \sigma )- 2c \Tr ( F_{\lambda \alpha
}F_{+\lambda})$. It is easy to
see that $\partial_\alpha J_\alpha =0$, as required by the consistency of
the equations. Since the
four-dimensional Laplacian is invertible, the above equation can easily be
solved, once we have
$J_\alpha$. For gauge group
$SU(2)$, $A_+$ is given by $$
A_+^a = \int d^4y~ \Delta^{ab}(x,y) \epsilon^{bkl} (A^k_\lambda \partial_+
A^l_\lambda
)(y)\eqno(11)
$$
where the Green's function $\Delta^{ab}(x,y)$ for $D_\lambda D_\lambda$ in
the instanton
background is given in reference [16]. To make the above solutions
explicit, we can, for example,
take the 't Hooft ansatz for instantons, viz., $A^a_\m = {\bar
\eta}^a_{\m\n} \partial_\n (\log \phi
)$ where $\phi = 1 + \sum_1^N {\rho_i^2 /
(x -a_i)^2} $ and insert it in various equations above. In this case,
$\sigma$, for example,
becomes $ 2c \partial_\m \phi \partial_\m \phi /\phi^2 $.

\noindent{\bf 4. String Interpretation.}

To see the stringy interpretation of our solution, we need to analyze its
moduli or zero mode
structure.
{}From the above equations, we see that, given the gauge field $F_{\m\n}$,
all the fields are
uniquely
determined up to the addition of the freely propagating six-dimensional
waves for the tensor
multiplet.
\footnote{*}{ Note the soliton
does not modify their propagation.}
Therefore the only zero modes correspond to the moduli of the instantons.

In order for our models to be mathematically meaningful they should be free
from local and global
gauge anomalies. In the absence of hypermatter, the gauge groups
$SU(2),~SU(3),~G_2,~F_4,~E_6,~E_7$ and $E_8$ can be made perturbatively
anomaly-free with the help of the Green-Schwarz prescription . However,
since the homotopy
group $\Pi_6$ of
the first three groups in this list are nontrivial, these theories will
harbour global gauge anomalies.
To
make them consistent we need to introduce hypermatter for these
theories[17]. The allowed matter
contents for the
cancellation of the global [17] as well as the local [18] anomalies in the
presence of one tensor
multiplet are
$n_2=4, 10$
for $SU(2)$, $n_3=0, 6, 12$ for $SU(3)$ and $n_7=1, 4, 7$ for $G_2$, where
$n_2, n_3$ and
$n_7$ represent the number of the doublets for $SU(2)$, triplets for
$SU(3)$ and 7-dimensional
representation of $G_7$, respectively. All other gauge groups are free from
global anomalies and
they can be made free from perturbative anomalies (using the Green-Schwarz
prescription)
if an appropriate amount of hypermultiplets are taken together with the
gauge and the tensor
multiplets
[17, 18].

For the gauge group
$SU(2)$, for the four-dimensional space being ${\bf R}^4$ and for instanton
number $k$,
we have $8k$ bosonic moduli corresponding to the instanton positions, scale
sizes and group
orientations.
(The equations of motion, despite the appearance of the dimensional
parameter $c$,
have scale invariance and give the scale size parameter in the solutions.)
These moduli appear
in the solution for the fields $B_{ab}$ as well.

The surviving supersymmetry has $\gamma_5 \varepsilon =\varepsilon$, i.e.,
left-chirality in the
four-dimensional sense corresponding to a $(4,0)$ world-sheet supersymmetry
for the solitonic
string. There must necessarily be fermionic zero modes. For the gauginos,
we have $4k$ zero
modes for the gauge group $SU(2)$, which are of right-chirality in the
four-dimensional sense and
are in the right-moving sector.
The Dirac equation for the gauginos along with the half-supersymmetry
condition shows that the
gaugino zero mode parameters are constants; the bosonic parameters are
constant as
well, by supersymmetry. The fermionic zero mode parameters are complex,
i.e., we have $8k$
real Grassman parameters which balance the $8k$ bosonic parameters. Some of
the fermionic zero
modes correspond to the supersymmetries which are broken by the background
and can be
obtained by such supersymmetry variations. With hypermatter, there are also
hyperino zero
modes, which are in the left-moving sector. There is no supersymmetry for
these modes and
generically there are no hyperscalar zero modes.

For higher gauge groups, there will be more moduli. Thus, for example, for
$SU(3)$, with the
standard
embedding of the instanton and $n_3=0$, we have $12k$ bosonic parameters
and $6k$ fermionic
parameters. It is easy to see that the number of moduli for all of the
anomaly-free gauge groups
listed above is always a multiple of $4$. We may thus interpret these
solutions as six-dimensional
strings
with $4$ transverse coordinates corresponding to the zero modes for the
broken translational
symmetries. The remaining zero modes can be regarded as additional
world-sheet degrees of
freedom. In this way for instanton number $k$, we have $k$ strings with
$(4,0)$ world-sheet
supersymmetry.

As an example, consider an $SU(2)$ theory with 10 hypermatter doublets
[19]. In this case,
for instanton number equal to one, we have eight instanton moduli, eight
gaugino zero modes for
the right-moving sector and 20 hypermatter zero modes for the left-moving
sector.
The $SU(2)$ symmetry can be spontaneously broken by vacuum expectation
values of
the scalars originating from the moduli corresponding to the global $SU(2)$
rotations and the scale
size of the instanton. By supersymmetry this should remove four of the
gaugino zero modes from
the right moving
sector by giving them a non zero mass, which will also eat up four hyperino
zero modes in the left
moving sector.
One is left with four moduli for
the instanton, four gaugino modes in the right-moving sector and 16
hyperino zero
modes in the left-moving sector. These 16 hyperino zero modes presumably
generate a left moving
$E_8$ current algebra. This looks like the spectrum of the non critical
string which lives in the
boundary of a membrane joining
a $5$-brane to a $9$-brane in $M$-theory and which becomes tensionless as
the $5$-brane
approaches the $9$-brane [20]. It has been argued in [21] that the same
model corresponds to one
of the phases of the $F$-theory.

There are also independent solutions with the opposite chirality. The
choice $\gamma_5 \varepsilon
=- \varepsilon$ leads to anti-self dual $H_{+\m\n},~F_{\m\n}$ with $A_+=0$
and $\partial_+ \xi
=0$.

The solution we have
obtained is a static one.
The choice of four-dimensional chirality as $\gamma_5 \epsilon =\pm
\epsilon$ leads to static
solitons.
By Lorentz boosts, it is possible to obtain a solution whose center of mass
is moving at a constant
velocity. For a moving soliton, the condition $\gamma_5 \epsilon =\pm
\epsilon$ must be
modified.
Consider, for example, the one-soliton (one-instanton) solution. We choose
the supersymmetry
parameters $\varepsilon$ as $S\varepsilon_{(0)}$ where $S=\exp ( -\half
\omega^\mu
\gamma_\mu) \approx 1 -\half \omega^\mu \gamma_\mu$ and $\varepsilon_{(0)}$
obeys
$\gamma_5
\varepsilon_{(0)}=\varepsilon_{(0)}$. (For small velocities, the parameter
$\omega^\mu \approx
v^\mu$, the velocity.) The vanishing of the gaugino variation, viz.,
$\Gamma^{ab}F_{ab}\epsilon =0$, now gives, to first order in $v^\mu$, $$
\eqalign{
F_{\mu\nu} - {\tilde F}_{\mu\nu}&=0\cr
F_{-\mu}+ {1\over {\sqrt {2}}}F_{\mu\nu}v^\nu &=0\cr F_{+-} - {1\over
{\sqrt{2}}}
F_{+\nu}v^\nu &=0\cr }\eqno(12)
$$
To this order, $F_{\mu\nu}$ is still self dual. The other two equations are
seen to be satisfied if
we
take the instanton position $a^\alpha$ to move with velocity $v^\alpha$,
i.e., $\partial_0 a^\alpha
=v^\alpha$. (We can make a gauge transformation $A_{-}\rightarrow A_{-}-
(1/{\sqrt{2}})A_\mu v^\mu $ to restore the $A_{-}=0$ gauge.) There is a
similar set of statements
for the vanishing of the tensorino variation. What we have shown is that a
soliton whose center of
mass is moving at a constant
velocity $v^\alpha$ is also a supersymmetric solution with supersymmetry
parameters being
$S\varepsilon_{(0)},~\varepsilon_{(0)}$ having definite four-dimensional
chirality.

\eject

\noindent{\bf 5. $D=6$ Models with Extended Supersymmetries.}

In the foregoing sections we discussed only the $D=6$ models with a minimum
number of
supersymmetries. Apart from the $(1,0)$ type supersymmetry in $D=6$ there
are also models with
$(1,1)$, $(2,0)$ and $(2,2)$ type supersymmetries. The number of real
components of the
supercharges are respectively $16$, $16$ and $32$. Out of these three types
only the $(2,0)$
models are chiral and therefore can be anomalous.
Like the four dimensional theories, if we do not want to have a physical
field of spin larger than 2,
then the total number of real supercharges should not exceed 32. This is
the number of
supersymmetries of the $D=11$ supergravity which is conjectured to be the
low energy limit of a
unifying theory of all known $D=10$ string theories and is called the M
theory. When we obtain a
lower dimensional theory from the $D=10$ or $D=11$ some of the super
symmetries can be
broken. For example the $K_3$ compcatification
which takes us from $D=10$ to $D=6$ breaks $1/2$ of supersymmetries. Thus
starting
from the type IIB theory, which has $32$ real chiral supersymmetries in
$D=10$, and
compactifying on a $K_3$ we obtain a $D=6$ theory with $16$ chiral
supersymmetries. This is an
example of a $(2,0)$ model with $21$ tensor multiplets [22]. It is exactly
$21$ tensor multiplets
which is required by the anomaly cancellation in $D=6$.
Although an invariant Lagrangian has not yet been constructed for these
models, the field equations
with an arbitrary number of tensor multiplets have been kn
own for some time [23].

The $(2,0)$ models involve self dual and anti-self dual tensor fields. One
can then
contemplate self dual or anti-self dual string like solutions of the type
discussed in the previous
section. Presumably these strings are also tensionless. An intuitive way of
understanding
this is
to remember that in $D=10$ the IIB theory has a four form potential whose
field strength is self
dual. There are also self dual $3$-brane solutions [24]. One can
imagine that a self dual brane wraps around a $2$ cycle of the $K_3$ to
produce an object which
will look like a string from the $D=6$ point of view. In general this
string will have some
thickness. But as the area of the cycle shrinks to zero the thickness also
will decrease.
Furthermore,
the string will be self dual by construction and its tension will be
proportional to the area of the $2$
cycle and hence will vanish as the area goes to zero.

The type IIA theory has the same
number of supersymmetries in $D=10$ as the type IIB, however it is a non
chiral theory. For that
reason upon compactification from $D=10$ to $D=6$ on a $K_3$ one obtains a
non chiral theory
in $D=6$ with $16$ real supercharges which generate a $(1,1)$
supersymmetry. The same type of
model can be obtained from the compactification of the heterotic strings on
a $T^4$. There exist
many compelling evidence that the theories obtained from the type IIA
compactifications on a
$K_3$ and the $E_8\times E_8$ compactifications on $T^4$ are dual in the
sense that the strong
coupling limit of one can be set in correspondence with the weak coupling
limit of the other [25].
At a first glance this looks puzzling, because, although at a generic point
on the moduli space of
the heterotic compactification the six dimensional gauge group is
$U(1)^{24}$ it is known from
Narain's work that at some special points the gauge symmetry can be
enlarged to a non Abelian
group [26]. For the duality to work one needs to find mechanisms for the
generation
of non Abelian gauge symmetries on the type IIA side. The possibility which
has been suggested
[27] is that the $2$-branes of the type II can wrap around $2$ cycles of
$K_3$ and produce, in
the limit that the area of the cycles shrink to zero, particle like objects
in $D=6$. The masses of
these
particles will be proportional to the area of the $2$ cycles and will
vanish in the limit of the
vanishing cycles. These massless particles should then match with the
massless particles generated at the special points in the heterotic moduli
space at which the gauge
symmetries are enhanced.

Finally the compactifications of type IIA or the type IIB theories on a
four dimensional torus will
produce non chiral $D=6$ supergravity models with $(2,2)$ supersymmetries.

An important role in the study of these models is played by the moduli
space of vacua, i.e. the
expectation values of the massless scalars. It has been conjectured that
all known compactifications
with $(1,1)$, $(0,2)$ and $(2,2)$ supersymmetries belong to the same moduli
space of vacua.

For example, we mentioned above that the heterotic string on $T^4$ is dual
to the type IIA
compactification on $K_3$. It is also known that if we compactify type IIA
on a $S^1$ of radius
$R$ from $D=10$ to $D=9$ then it gives the same theory as the one obtained
from the
compactification of type IIB from $D=10$ to $D=9$ on a circle of radius
$1/R$. It then follows
that the compactification of the heterotic string on a $T^5$ should produce a
theory dual to the compactification of the type IIB on $S^1\times K_3$.
This duality is a strong
weak duality. On the heterotic side
the string coupling is given by $e^{\phi}$, where $\phi$ denotes the vev of
the dilaton field. On
the type IIB side this modulus corresponds to the radius of $S^1$. This
implies that the strong
coupling limit of the
heterotic string in $D=5$ is dual to the large radius limit of the type IIB
string in $D=5$. The large
radius limit is, of course the same as the decompactification limit which
takes us back to $D=6$
space time again. Thus by moving around in the moduli space of type IIB or
heterotic in $D=5$ we
can end up with the $(2,0)$ type IIB in $D=6$.

\vskip .2in

\vskip .2in
\noindent{\bf References}
\item{1.} S. Randjbar-Daemi, Abdus Salam and J. Strathdee, Nucl. Phys. {\bf
B214} (1983)
491.

\item{2.} For excellent reviews see, for example, the contributions by A.
Sen, P. K. Townsend
and C. Vafa in the Proceedings of the 1996 Trieste Summer School.

\item{3.} For a review of the electromagnetic duality and the Olive
Mantonen conjecture see D.
Olive in ICTP Conference on Recent Developments in Statistical Mechanics
and Quantum Field
Theory, Trieste, Italy, 10-12 Apr 1995.
Published in Nucl. Phys. Proc. Suppl. 45A:88-102, 1996, hep-th/9508089.

\item{4.} M.J. Duff and R. Minasian, Nucl. Phys. {\bf B436} (1995) 507.

\item{5.} P. K.
Townsend, Phys. Lett. {\bf B202} (1988) 53.

\item{6.} S. Randjbar-Daemi, Abdus Salam, E. Sezgin and J. Strathdee, Phys.
Lett.
{\bf B151} (1985) 351.

\item{7.} M. Green, J. Schwarz and P. West, Nucl. Phys. {\bf B254} (1985) 327;
J. Schwarz, Phys. Lett. {\bf B371} (1996) 223, hep-th 9512053.

\item{8.} M. Green and J.
Schwarz, Phys. Lett. {\bf B149} (1984) 117.

\item{9.} B. Zumino, in ``Relativity, Groups and Toplogy II", Eds. B.S. De
Witt and R. Stora
(North-Holland, 1984).

\item{10.} A. Sagnotti, Phys. Lett. {\bf B294} (1992) 196, hep-th/9210127;
Fabio Riccioni and
Augusto Sagnotti, hep-th/9806129.

\item{11.} E. Witten, Phys. Lett. {\bf B117} (1982) 324; S. Elitzur and
V.P. Nair, Nucl. Phys.
{\bf B243} (1984) 205.

\item{12.} N. Seiberg and E. Witten, hep-th/9603003; N. Seiberg, {\bf B
390} (1997) 169, hep-
th
9609161.

\item{13.} E. Bergshoeff, E. Sezgin and E. Sokatchev, Class. Quantum Grav.
{\bf }

\item{14.} V.P. Nair and S. Randjbar-Daemi, Phys. Lett. B418:312-316,1998
hep-th/9711125.

\item{15.}M.J. Duff, J.T. Liu, H. Lu, C.N. Pope, hep-th hep-th/9711089.

\item{16.} L.S. Brown {\it et al}, Phys. Rev. {\bf D17} (1978) 1583.

\item{17.} M. Bershadsky
and C. Vafa, hep-th 97030167.

\item{18.} U.H. Danielsson, G. Ferretti, J. Kalkkinen and P. Stjernberg,
hep-th 9703098.

\item{19.}P. Mayr and C. Vafa, unpublished.

\item{20.}Ori J. Ganor and Amihay Hanany, hep-th 96020120.

\item{21.} A. Klemm, P. Mayr and C. Vafa, hep-th 96070139.

\item{22.} P.
Townsend, Phys. Lett. {\bf B139} (1984) 283.

\item{23.} L. Romans, Nucl. Phys. {\bf
B276} (1986) 71.

\item{24.}M.J. Duff and J.X. Lu, Phys. Lett. {\bf B273} (1991) 41.

\item{25.} P.
Townsend and C. Hull, Nucl. Phys. {\bf B243} (1984) 205, hep-th/9410167.

\item{26.} K.S. Narain, Phys. Lett. {\bf B169} (1986) 409; K.S. Narain,
M.H. Sarmadi and E.
Witten, Nucl. Phys. {\bf B279} (1987) 369.

\item{27.} E. Witten, Nucl. Phys. {\bf B443} (1995) 85, hep-th/9503124;
P.S. Aspinwall, Phys.
Lett. {\bf B357} (1995) 329, hep-th/9507012.

\end